\documentclass{aa}

\usepackage{graphicx}
\usepackage{epsfig}

\titlerunning{The Outer Atmosphere of Betelgeuse}

\begin{document}

\title{The Outer Atmosphere of the M-type Supergiant $\alpha$~Orionis: K\,{\sc i} 7699\AA\ Emission}

\author{Bertrand Plez \inst{1}
\and David L. Lambert \inst{2}}

\institute{GRAAL Universit\'{e} de Montpellier II, F-34095 Montpellier-Cedex 5, France
\and Department of Astronomy, University of Texas at Austin, Austin TX
78712-1083}

\offprints{B. Plez,
plez@graal.univ-montp2.fr}

\date{Received <date> / Accepted <date>}

\abstract{
Spatially-resolved high-resolution long-slit
spectra of Betelgeuse's circumstellar
shell are described for a spectral window centered on the 7699\AA\
resonance line of neutral potassium. The K\,{\sc i} emission from
resonance fluorescent scattering of photospheric photons which is mapped
out to 50 arcsec from the star is approximately
spherically symmetric with a brightness decreasing as 
$r^{-2.36 \pm 0.03}$,
where $r$ is the radial distance from the star. Our measurements together with
the earlier theoretical interpretation by Rodgers \& Glassgold suggest that the
mass loss rate is about 2 $\times 10^{-6}$ $M_\odot$ y$^{-1}$.
The K\,{\sc i} emission is far from homogeneous: intensity inhomogeneities
 are seen down to  the seeing limit of about 1 arcsec
and the velocity resolution of about 2 km s$^{-1}$.
There is clear evidence for a thin shell of 50  arcsec
radius. This is identified with the weaker circumstellar absorption
component known as S2. Estimates are made of the density of K atoms in
this shell ($\simeq 6 \times 10^{-5}$ cm$^{-3}$).
}

\maketitle


\section{Introduction}

Massive stars may evolve to become red supergiants prior to death by 
supernova explosion. Since mass lost by the red supergiant may dictate
the time and form of the explosion and also the nucleosynthetic
 contribution of the massive
star to the chemical evolution of the Galaxy, studies of mass loss by
red supergiants are of great interest. Mass loss rates are not yet calculable
from an {\it ab initio} theory.  
Although measurements of the
mass loss rates are available from a variety of observations of the gas
and dust in the circumstellar winds, uncertainties remain large. One factor
that is often a contributor to the uncertainties  is  an
assumption that the stellar surface, 
 circumstellar shell and the wind 
are spherically
symmetric. Observations are beginning to reveal that supergiants
do not recognize fully this simplifying assumption.

Direct observational tests of the assumption are most easily executed on
the nearest red supergiants.
 In the case of Betelgeuse ($\alpha$ Orionis), the
nearest of these stars, departures from spherical symmetry have been
extensively noted. At ultraviolet, visible, and near-infrared wavelengths,
observations have shown the presence of `bright spots' on or near the stellar
surface (cf. Gilliland \& Dupree 1996; Buscher et al. 1990; Young et al.
2000). Such spots that change with time
are generally attributed to hot rising convective
cells, as predicted by Schwarzschild (1975) and recently 
shown by Freytag (2002) in 3D hydrodynamical simulations (the hot
spots appear occasionally over downdrafts), but an alternative
origin has been proposed (Gray 2000). 
Comparison of ultraviolet and radio brightness 
temperatures led Lim et al. (1998)
to conclude that strong temperature inhomogeneities extend to at least
several stellar radii above the surface. The measured radio brightness
temperature at 6cm of about 1400K corresponding to a height of
about 7 stellar radii is in good agreement with the temperature of about
1200K derived by Justtanont et al. (1999) from infrared
 ground state fine structure
lines of Si\,{\sc ii} and Fe\,{\sc ii}. Temperatures derived from ultraviolet
spectra  are much higher at these heights implying that cold and warm
gas coexist at the same heights above the stellar surface (Lim et al. 1998).
Recently, Harper, Brown, \& Lim (2001) have constructed a semiempirical
model for the inner shell extending from the photosphere to about 40 stellar
radii (about 2 arcsec).
It is not unlikely that the inhomogeneities detected in ultraviolet and
radio observations persist to greater heights and into the stellar
wind.

Dust in the shell close to the star has been resolved by infrared
interferometry and direct imaging, and detected
far from the star  by the polarization that
scattering of dust grains introduces.  Infrared emission at about 11 $\mu$m
from silicate dust grains comes from an extended region of several
arcsec with an inner shell radius of 1 to 2  arcsec
(cf. Sloan, Grasdalen, \& LeVan 1993; Danchi et al. 1994; Bester et al. 1996; 
Rinehart, Hayward \& Houck 1998; Sudol et al. 1999). Undoubtedly,
dust  - cooler grains - extends to greater distances but has yet to
be resolved by far-IR imaging. Extended emission in IRAS images was
noted by Stencel, Pesce, \& Hagen (1988) and Young, Phillips, \& Knapp
 (1993a, 1993b).  
Warm and cold dust is observable by the photospheric light that
grains scatter. This scattered light is separable from light scattered
by the Earth's atmosphere and the telescope by its polarization: the
polarization vector is tangential to the radius vector. McMillan and
Tapia (1978) following a suggestion by Jura and Jacoby (1976) detected
the polarized scattered light from Betelgeuse's shell. Additional
measurements were made by Le Borgne, Mauron, and Leroy (1986) who
show that the number density of dust grains declines roughly as the
inverse-square of the radial distance from the star for distances of
10 to 90 arcsec.

Circumstellar gas as detected by blue-shifted absorption lines
was observed and analyzed by Weymann (1962) who provided 
indirect evidence  that the shell extended to great distances
from the star.
Direct evidence in the case of the visual binary $\alpha^1$ Herculis
for a very extended shell off this M supergiant was presented by Deutsch (1956)
who found circumstellar lines in the spectrum of the visual companion to the
supergiant. Weymann's inferences about the shell size were compatible with
the direct lower limit to the outer shell radius from Deutsch's observation.
Observations of H\,{\sc i} 21 cm emission from Betelgeuse have shown that gas
extends to about 1 minute of arc from the star (Bowers \& Knapp 1987).

Betelgeuse's circumstellar shell was  resolved spatially  by
Bernat \& Lambert (1975) who  detected fluorescent emission in the
K I 7699\AA\ resonance line at angular distances of 2 - 4 arcsec from
the star.
Subsequent observations extended the detections of the 7699\AA\ 
emission to about 60 arcsec from the star and provided evidence for
some departures from spherical symmetry and for structures in the shell
(Bernat \& Lambert 1976; Bernat et al. 1978; Honeycutt et al. 1980;
Mauron et al. 1984;  Mauron 1990). Emission in the Na D lines was
detected by Mauron \& Querci (1990), Mauron (1990), and Mauron \& Guilain
(1995). The most remote (from the star) manifestation of the
circumstellar shell is the parsec-sized bow shock reported by
Noriega-Crespo et al. (1997) from IRAS images at 60 and 100 $\mu$m. Emission
detected from an arc with a mean radius of 6 minutes of arc is attributed
to material confined by the interstellar medium's ram pressure.

In this paper, we present and analyse a new series of observations of
K\,{\sc i} 7699\AA\ emission obtained with a  long slit at a resolving power of
about 110,000  or 2.6 km s$^{-1}$ with the slit placed at
various impact parameters from the star and position angles around the
star, and at several epochs. Our combination of spectral and spatial
resolution was not achieved by earlier studies of the K\,{\sc i}
emission.
We derive afresh the radial
dependence of the K\,{\sc i} emission and rediscuss the principal
conclusion - the mass loss rate - that has been previously drawn from
the radial dependence. A novel result of our observations is the
discovery of clumps in the emission and, hence, in the neutral
gas in the shell and wind. Some of these inhomogeneities
are unresolved both spatially in our seeing-limited
observations and also kinematically at the limit of about 2 km s$^{-1}$.

\section{Observations and Reductions}

Betelgeuse was observed  in 1994 February, March, and
November with the W.J. McDonald Observatory's
 Harlan J. Smith 2.7m reflector  and its coud\'{e} spectrograph. The latter
was used with an echelle grating and an interference filter to separate
the order providing the K\,{\sc i} 7699\AA\ line. The slit length
was varied with a majority of the observations taken with a length of
10 to 100  arcsec. The slit width was almost always set at 0.55
 arcsec.  
Spectra were recorded by a Tektronix 512 X 512 CCD  over
a bandpass of about 17\AA.  Standard reduction procedures were applied
to correct for the bias and pixel-to-pixel variations. 
The wavelength scale was set using Th-Ar lamp spectra.

During each observing session, several on-star spectra were recorded with a very
short slit length.  Off-star spectra were then collected with longer
slit lengths. In general, the star was not included in the slit; inclusion
limited the exposure time beyond which the starlight saturated a portion of the
CCD chip. Since the field at the coud\'{e} focus rotates, exposure times
were kept short to minimize smearing of emission inhomogeneities;
typically, exposure times were 10 minutes or less. For capturing emission
far from the star, a few exposures as long as 30 minutes were made.
Relative spectrophotometry
was attempted;  observations were terminated in partly cloudy skies.
The stellar flux was measured by opening the slit to its maximum 
width when all the visible image was contained within the slit.
A guider maintained the star's position relative to the slit to an
accuracy of about 0.5  arcsec.

\begin{figure}
\centerline{\epsfig{figure=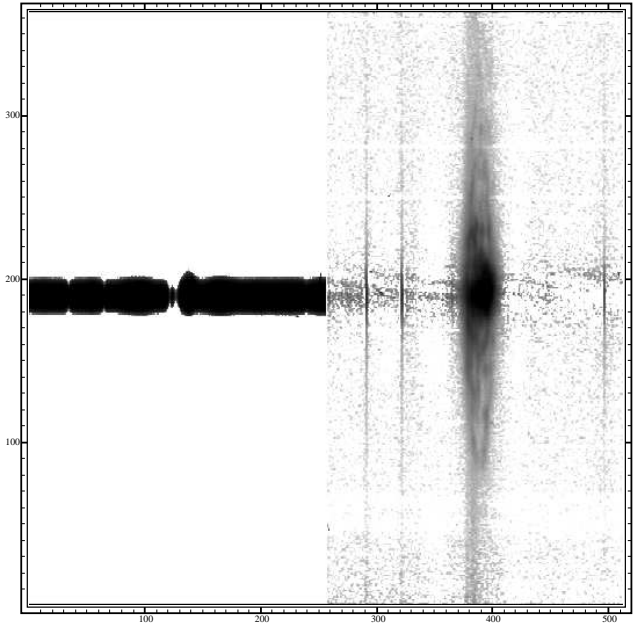,bbllx=25pt,bblly=170pt,bburx=555pt,bbury=695pt,clip=,width=8.5cm}}
\caption{Leftpanel: raw image. Dispersion along the horizontal axis 
(1.3km.s$^{-1}$/pixel). Blue to the left. The slit, positioned 3" away from 
the star,
 spans 100" on the vertical axis, with a scale of 0.27"/pixel. 
Right panel: Velocity-position map which is the raw image
 after subtraction of the stellar 
light scattered by the atmosphere and the instrument. Pixels are numbered
along both axes.}
\label{fig1}
\end{figure}

Our reduction techniques were adapted directly from those described
earlier (Plez \& Lambert 1994). In short, the spectrum at a given location
off the star is presumed to be a composite of emission from the gas down
the line of sight through the shell at that point and starlight
scatter in the Earth's atmosphere and the telescope. We assume that the
latter contribution may be represented by a scaled version of an on-star
spectrum. Fig.~\ref{fig1}  
provides an example of a raw off-star spectrum and
the spectral image after subtraction of the photospheric scattered
light. 
For positions of the slit near 
the star (a few arcsec), the photospheric scattered contribution 
largely dominates over the light scattered by the shell. This makes
the subtraction difficult, as the imperfectly removed background
spectrum shows in Fig.~\ref{fig1}. The terrestrial O$_2$ lines 
appear weakly in emission in the image after subtraction. This is due to 
the difference in illumination across the slit when the star is
 on one side of the
slit. For spectra taken very near the star, we could improve  subtraction
of the photospheric component by shifting the spectra by less than
half a pixel in the dispersion direction.
Slight changes in airmass between the on-star and off-star exposures
may also affect the subtraction of the O$_2$ lines.

The spectrum after subtraction of the scattered light is a velocity-position
map of the emission. Rows of pixels parallel to the dispersion give the
spectrum at different points in the shell. The radial distance of a point from
Betelgeuse may be calculated from the impact parameter (3  arcsec in this
case) and the distance in seconds of arc from the point on the slit of
closest approach to the star. The orientation of the slit was derived from
the hour angle of the observation.  A thin uniform shell of outwardly
flowing and
fluorescing K\,{\sc i} atoms appears as an ellipse in such a velocity-position
map. Such an elliptical looking feature is evident in Fig.~\ref{fig1}. The extent
of the ellipse along the slit is set by the radius of the shell and the
impact parameter. The maximum width of the ellipse is directly related to the
expansion velocity and  impact parameter; the width vanishes when the impact
parameter equals the shell's radius.

Several assumptions underlie our reductions.
The recorded off-star spectrum results from  a convolution of the 
light from points around a position on the slit with the seeing profile. 
A deconvolution of the observed spectra to obtain the true
spectrum at a point is impossible given the varying seeing conditions.
We assume that the dominant contaminant is scattered starlight with a spectrum
represented by the on-star spectrum.
 The on-star spectrum
is itself, however,  not devoid of signatures of
 circumstellar K\,{\sc i} absorption and
emission at 7699\AA. In particular, there is  strong blue-shifted
absorption 
introduced by the potassium atoms along the line of sight to the star.
At some level, this absorption is weakened by emission from the immediate
environs of the star. Certainly, the angular diameter of the stellar
photosphere of about 0.05  arcsec is considerably smaller than
the smallest aperture (0.55 x 0.55 
arcsec) used for the on-star
spectra. There is no obvious way to retrieve the true stellar spectrum from
the data. A predicted photospheric profile could be used but this would
not account for the fact that much of the shell absorption probably
occurs close to the star.
In view of the low inferred  intensity (relative to the stellar scattered
light) and the steep gradient
 of the emitted light from the shell, we suggest that our simple reduction
technique is adequate.

Our assumption that the circumstellar shell's contribution is solely
emission in the K\,{\sc i} line ignores a contribution from the
dust grains that scatter photospheric light. This scattered light
from different locations in the shell will be Doppler-shifted
by different amounts off the (assumed) radially-flowing dust grains, i.e.,
the scattered light will be a smeared version of the photospheric
spectrum.
 Optical depth due to dust is small and multiple scattering of
photons may be neglected.   Circumstellar gas is flowing out at a
velocity of about 10 km s$^{-1}$. If the dust and gas are well coupled,
the Doppler shifts imposed on light scattered by dust grains will
amount to a line broadening of less than about $\pm$10 km s$^{-1}$. Since the
intrinsic widths of the photospheric lines exceed this value, it will
be difficult to distinguish the dust-scattered contribution from that
scattered by our atmosphere and telescope. Such a distinction is
especially hard to achieve with our spectra in which the K\,{\sc i}
line is the sole strong line. Perhaps, the isolation of the dust-scattered
component can be made using a region richer in deep absorption lines.
It may be necessary to exploit the polarization of the scattered light
to effect a clear isolation of the dust-scattered light.
We assume that the dust  makes a negligible contribution to the
total scattered light.

Our interpretations are based on the assumption that the K\,{\sc i}
emission results from fluorescence and not from radiative recombination
or collisional excitation. Calculations using densities and temperatures
expected of the stellar wind fully support this assumption.

\section{An Asymmetric and Clumpy Shell}

\begin{figure}
\centerline{\epsfig{figure=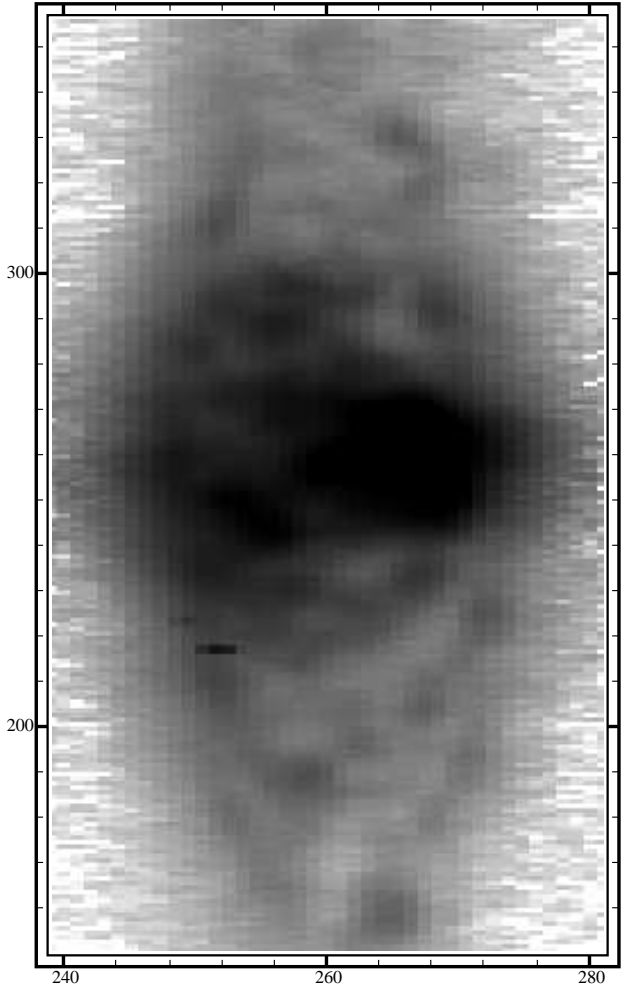,bbllx=150pt,bblly=210pt,bburx=433pt,bbury=655pt,clip=,width=8.5cm}}
\caption{Velocity-position map. Slit positioned 4" away from the star.
Dispersion runs horizontally and is 1.3km.s$^{-1}$/pixel. The slit spans 
56" with a pixel size of 0.27". The star position is on line 253. 
Note the very rich and clumpy structure, 
revealing successively larger incomplete emission
shells. At this position
on the sky, around column 260,  there is more red-shifted than blue-shifted emission.
Pixels are numbered along both axes.}
 \label{fig2}
\end{figure}

\begin{figure}
\centerline{\epsfig{figure=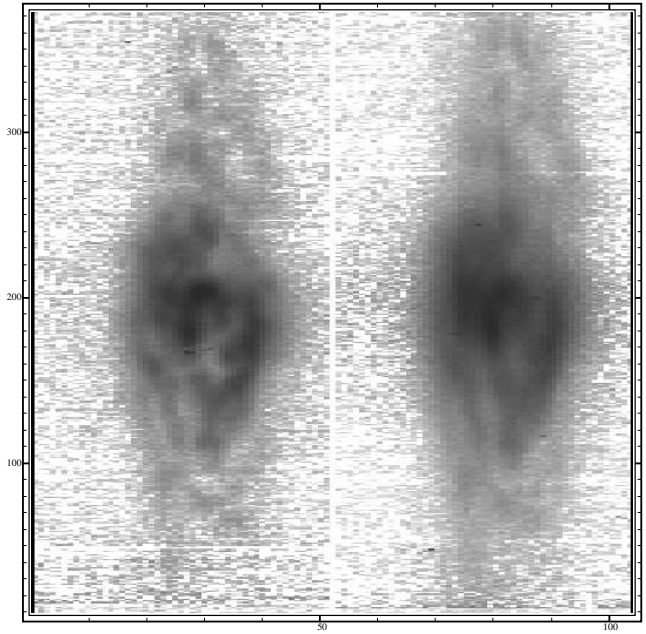,bbllx=29pt,bblly=170pt,bburx=552pt,bbury=695pt,clip=,width=8.5cm}}
\caption{Two observations at the same orientation and
 slit position (13" 
 from Betelgeuse).
Velocity scale as in Fig.~\ref{fig2}: 0.27"/pixel along the slit(vertical axis), 
and 1.3km.s$^{-1}$/pixel on the dispersion axis (horizontal). The star position
corresponds to  line 190. Left panel:
March 22, 1994 with 1.2" seeing. Right panel: Nov 28, 1994, with seeing around 3".
Note the great similarity of the images. Most differences are probably due
to seeing degradation. Pixels are numbered along both axes.}
 \label{fig3}
\end{figure}

\begin{figure}
\centerline{\epsfig{figure=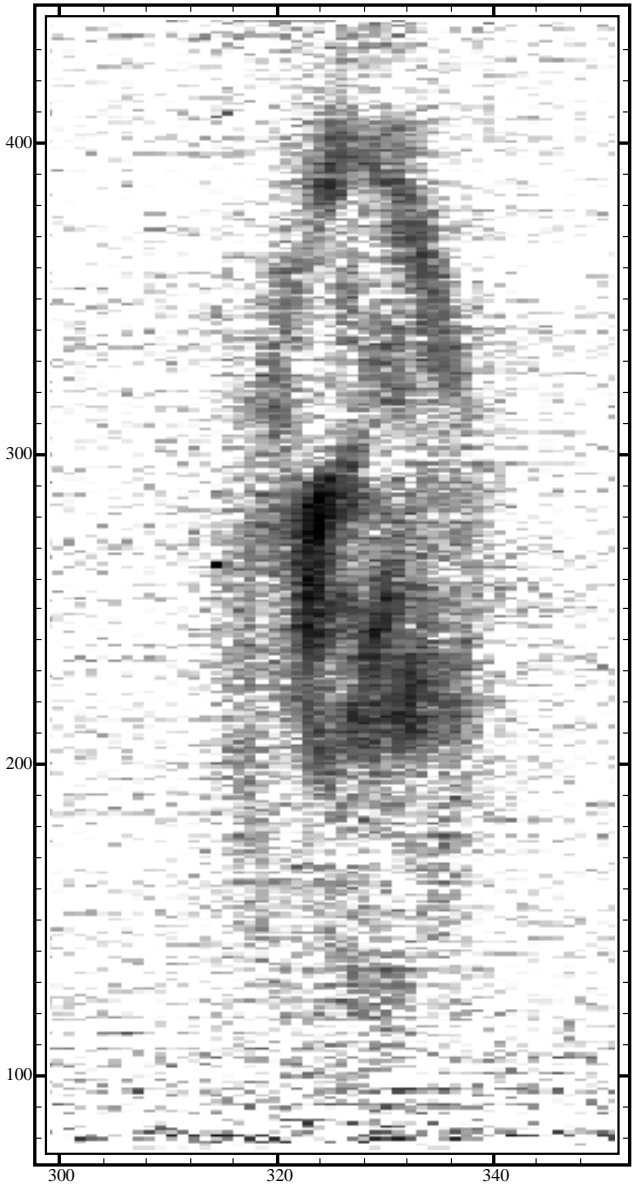,bbllx=148pt,bblly=162pt,bburx=436pt,bbury=703pt,clip=,width=8.5cm}}
\caption{The S2 shell. Slit at 33" from the star. Scale as in
Fig.~\ref{fig3}. Star position corresponds to line 261. Pixels are numbered along both axes.}
 \label{fig4}
\end{figure}

Inspection of the reduced spectra shows that the circumstellar
envelope as traced by the K\,{\sc i} emission about the star is not
completely spherically symmetric. 
A particularly striking feature
of the velocity-position maps is the appearance of clumps at
scales extending down to the spatial and spectral resolution of
our observations. This degree of clumpiness has not been revealed by
previous observations.
Sample velocity-position
maps are shown in Fig.~\ref{fig2}, \ref{fig3} and \ref{fig4} 
for slit positions at increasingly larger distances from the star illustrate
these departures from spherical symmetry which we discuss in
detail below.
These Figures also show that the maximum velocity extent 
of a  velocity-position map is not greatly different for small and 
large impact parameters.

\subsection{Radial Dependence of the K\,{\sc i} Emission and Mass Loss Rate}\label{radial}

\begin{figure}
\centerline{\epsfig{figure=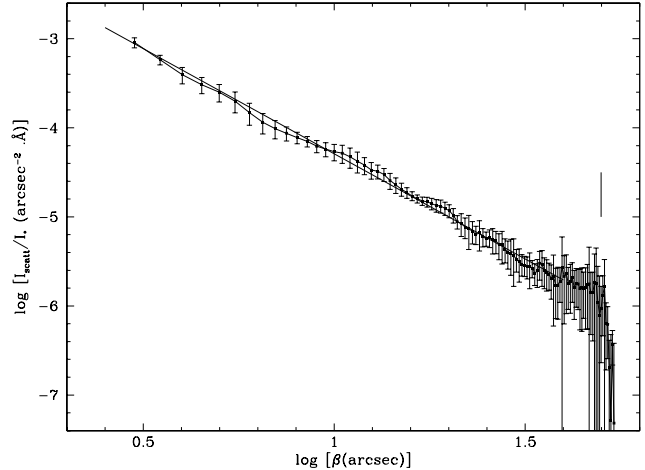,bbllx=41pt,bblly=27pt,bburx=578pt,bbury=749pt,clip=,angle=-90,width=8.5cm}}
\caption{Average scattered intensity profile. 
I$_{scatt}$/I$_{*}$ is the
intensity scattered by the shell, normalized to the photospheric flux (see text
for details). Eleven profiles from exposures at various distances from the
star were averaged. In each individual image, the intensity was integrated
in 1" bins along the slit. Error bars are one standard deviation. The best linear
fit to the data out to 50 arcsec is superimposed (slope=-2.36). The position of
 the 50 arcsec shell is indicated by the tick mark. Note the strong drop in intensity
beyond that distance to the star.}
 \label{fig6}
\end{figure}

By azimuthal averaging to the maximum extent possible for a given sequence
of observations, it is possible to determine the mean radial
dependence of the emission intensity integrated over the line profile. This was
done only for exposures obtained under good photometric conditions. The
intensity of the scattered light at the radial distance $r$ is
written as $I_{scatt}(r)$. We scale the intensity to the measured stellar
flux $I_*$ measured with the very wide slit. $I_*$ was measured by integrating
the on-star spectrum flux over the whole spectral interval 
(512 pixels corresponding to 17.24\AA). This total flux was then divided
by the wavelength range and  the 
integration time. The resulting $I_*$ is thus
expressed in counts per \AA\  per second.
Note that Mauron et al. (1984) used a similar definition
in  averaging the flux over
20\AA, while Mauron (1990), Honeycutt et al. (1980), and Bernat et al. (1978)
used the photospheric flux in 1\AA\ centered on 7699\AA.
There are a number of  sources of uncertainty affecting the
determination of the photospheric flux and
in the $I_{scatt}(r)/I_*$ ratio, 
 mostly
because of possible loss of photons intercepted by the slit jaws, and
changes of transparency and seeing conditions between exposures.
 The derived normalized
intensity of the K\,{\sc i} emission, $I_{scatt}(r)/I_*$, is shown in
Fig.~\ref{fig6} 
 for impact parameters $R$ out to 50  arcsec from the star.
The mean profile is consistent with a power law of slope -2.36 $\pm$ 0.03
(least-square fit to the data).
 Perhaps, one should fit the data with a two component model:
a slope of -2.59 $\pm$ 0.2 out to 7  arcsec and  of 2.55 $\pm$ 0.07 
beyond 10 arcsec but brighter by a factor of 1.4. 

Our measured slope is equal within the uncertainties to that
reported earlier by Honeycutt et al. (1980) who gave the slope as
-2.65 $\pm$ 0.2 from average measurements in the four principal
quadrants at radial distances of about 10 to 50  arcsec. 
Estimates of the
intensity of the scattered light are similarly in good agreement:
we obtain $I_{scatt}/I_* \simeq 10^{-5.0}$ at 20 arcsec from the
star, and Honeycutt et al. reported a value of $10^{-4.8}$.  The agreement is
less satisfactory with Mauron  et al. (1984) who reported a slope of
-3.5 $\pm$0.8 from measurements spanning distances of 20 to 55 
arcsec but the intensity at 20  arcsec ($10^{-4.8}$) is consistent
with other measurements. Mauron's (1990 - see also Mauron \& Querci 1990)  
measurements at 5 to 8  arcsec are similarly a little higher in
intensity than our mean values. Although the differences could be due
to changes in the efficiency of fluorescence, the density of potassium
atoms in the shell, and other physical factors, we think it more likely that
they reflect observational errors encurred in the surface photometry.
As noted by Mauron (1990), earlier results by Bernat et al. (1978) gave
much higher intensities for the emission close to the star than are now
reported: e.g., Bernat et al. found $I_{scatt}/I_* \simeq 10^{-1.3}$
at about 3 arcsec but our value is 10$^{-3.1}$. This large difference
is unlikely to be due to changing conditions at the stellar surface or in
the shell but reflects the difficulty of making photometric measurements of
a  faint
light source close to a bright star with a  detector (a vidicon) with a
non-linear response.

A calculation of the stellar mass loss rate from the radial dependence
of the 7699\AA\ emission has been discussed previously, notably by
Rodgers \& Glassgold (1991).
 Although the
observations may provide reliable estimates of the run of the density of
potassium atoms, an estimate of the mass loss rate  calls for knowledge 
the degree of ionization of potassium, the depletion of potassium
onto dust grains,  excitation of the atoms
(fluorescence vs radiative recombination, for example) in the shell,
and radiative
transfer  through the
shell. We assume the shell to be optically thin to the resonance photons,
a conclusion that is not contradicted
 by the ratio of emission in the two resonance
lines (Mauron \& Querci 1990). Here, we adopt
the recipes provided by Rodgers \& Glassgold and adjust their derived
mass loss rate to our measured intensities.

The slope of the   $I_{scatt}(r)/I_*$ versus $r$ relation is, if
spherical symmetry is assumed, directly related to the radial
dependence of the number density of neutral potassium atoms. Additional
assumptions include uniform illumination of the atoms by the photosphere,
and a shell that is optically thin to the 7699\AA\ photons. It is then
readily shown that a relation $N(r) \propto r^{-\theta}$ provides an
intensity dependence $I_{scatt}/I_* \propto  r^{-(\theta + 1)}$ (Bernat 1976;
Gustafsson et al. 1997). If  neutral potassium atoms  are
conserved in a radial outflow, their  density dependence follows $\theta = 2$
and $I_{scatt}/I_*$ follows a power law of slope 3. Conversely, the
observed slope of -2.36 indicates a density dependence with a slope
of  -1.36.
 Our observed slope is in good agreement with the value of
-2.33 corresponding to the model by Rodgers \& Glassgold. 
The difference between their prediction and the slope of -3
predicted for a uniform shell with conservation of potassium atoms 
is due to their introduction of ionization of 
potassium atoms in the shell by chromospheric ultraviolet photons. 
Glassgold \& Huggins (1986) and Rodgers \& Glassgold (1991)
discuss in detail the run of the neutral
potassium fraction. 
The ionizing chromospheric radiation field
is not much shielded in the envelope and decreases as $r^{-2}$.
The electron fraction, set constant in Glassgold \& Huggins model, 
increases slightly outwards in Rodgers \& Glassgold's more detailed model 
resulting in a gradient
of the electron density which is shallower than
$r^{-2}$. The
recombination coefficient increases with distance as a result
of the temperature decrease. The neutral fraction thus increases with 
distance from the star in the inner shell.
At large distances out from the
star  where  ionization of potassium atoms is controlled by the interstellar
radiation field, the slope of $I_{scatt}(r)/I_*$ must
steepen. This steepening is visible beyond
about 30~arcsec in our data on Fig.~\ref{fig6},
although at first sight masked by the detached shell S2 (see Sec.~\ref{detached}).
The slope increase (to about -4.2) is consistent with Rogers and Glassgold (1991) model.

Following Gustafsson et al. (1997) we write:

$${I_{scatt}(\beta)\over I_*}={\pi e^2\over m_ec^2} \lambda^2 10^8 f {d^2 \over 4\pi 206265^2}
\int_{-\infty}^{+\infty}{N_{scatt} {dz\over r^2}} {\rm arcsec^{-2}.\AA},$$

where $\beta$ is the angular distance from the star in arcsec, $f$
is the oscillator strength of the line, $N_{scatt}$ is the number density of 
scattering atoms in cm$^{-3}$, $d$ is the distance to the star in parsec, 
and $z$ is the distance along the line of sight in cm.
Assuming a power law of the form: $N_{scatt} = N_{s15} ({10^{15}\over r})^{\theta}$,
with $N_{s15}$ the number density at $r=10^{15}$cm, the integral is readily
calculated and yields: 
$N_{s15} (10^{15})^{\theta} Y ({206265\over \beta d})^{(\theta +1)}$,
with $Y$ a constant depending on the value of $\theta$.
From our observations we derive 
${I_{scatt}(\beta)\over I_*}={0.0117\pm 0.0011\over \beta^{2.36\pm 0.03}}$.
Adopting a distance of 140 pc for Betelgeuse from its 
Hipparcos parallax of $7.63\pm1.64$~mas, we find the number 
density of scattering atoms to be
$N$(K\,{\sc i}) = 6.0 $\times 10^{-4} \times (10^{15}/r)^{1.36}$ cm$^{-3}$.\footnote{Gray 
 (2000) has remarked that the accuracy of the Hipparcos parallax
may be seriously underestimated because it assumes that the centroid of the
stellar image remained fixed relative to the stellar disk over the
period of observations; a bright spot midway between disk center and the
stellar limb and contributing 25\% of the light could cause the centroid
to shift by half the parallax.} 

The column density of neutral K atoms along the line of sight to the
star was derived by Bernat (1977) from the 7699 \AA\ circumstellar
absorption line: $N$(K\,{\sc i}) $\sim 4 \times 10^{12}$ cm$^{-2}$. Integrating
our relation for the radial dependence, Bernat's column density is
reached by integrating from infinity to an inner radius $R_{inner}$
= $9 \times 10^{13}cm$ $\simeq 2R_{star}$. This is a plausible value for the
inner radius. To halve the column density requires
$R_{inner} \simeq 13R_{star}$. To double for $R_{inner} < R_{star}$, 
which makes no sense. In short,
our derived radial dependence for the density of potassium atoms is
consistent with the column density derived from the absorption
line. 

For this reason and because
our observed slope for $I_{scatt}(r)/I_*$ is equal to that
predicted by Rodgers \& Glassgold, we use the  ratio of our
observed to their predicted number
density versus radial distance directly to correct their 
mass loss rate.
This corresponds to a reduction of their mass loss rate from $\dot{M}$ =
 $4 \times 10^{-6}$
$M_\odot$ y$^{-1}$ to 
$2.4 \times 10^{-6}$ $M_\odot$ y$^{-1}$.  The revision is dependent
on
our value of $I_{scatt}/I_*$ which
 rests on an average of the stellar photospheric
flux averaged over 17 \AA\ around the potassium line. The stellar flux
intercepted by the outflowing gas of the shell is that on the blue wing
of the K\,{\sc i} line. As already pointed out, we do not have access 
to the true photospheric spectrum, and the K\,{\sc i} line profile can only
be guessed at. A rather conservative estimate of a factor of 2 lower 
flux at line center leads to a  mass-loss rate larger
by a factor of two. A distance of 200 pc was assumed by Rodgers \& Glassgold
and Glassgold \& Huggins. To correct the mass loss rate to the Hipparcos
distance of $d$ = 140 pc,
 we use the scaling given by Mauron \& Caux (1992), i.e.,
the K\,{\sc i} brightness at a given angular distance
from the star  is approximately proportional to $\dot{M}^2/d$. Adoption of
 $d$ = 140 pc instead of 200 pc implies a reduction of $\dot{M}$ by 85 \%
or $\dot{M} = 2 \times 10^{-6} M_\odot$ y$^{-1}$. 
Glassgold \& Huggins warn that the accuracy of an $\dot{M}$ derived from
K\,{\sc i} emission is low (``a factor of some 3 - 5''). Although the
subsequent analysis of the temperature profile (Rodgers \& Glassgold 1991)
may have reduced the uncertainty, a factor of two uncertainty seems
likely to be optimistic. The derived rate is dependent too on the
assumption of a spherically symmetric shell, an assumption probed
observationally in the next section.

This uncertainty could in principle be reduced by observing other
emission lines. 
In light of the higher abundance of Na and using the atomic concentrations
predicted by
Rodgers \& Glassgold's (1991), the Na D emission is expected to be
stronger than that in the K\,{\sc i} 7699 and 7665 \AA\ lines. Mauron \&
Guilain (1995) find that the Na D emission is considerably fainter
than expected although other mass losing M stars do show Na D emission
stronger than the K\,{\sc i} emission. These authors suggest that
interstellar absorption is ``the simplest explanation, though not completely
convincing''. Fortunately, interstellar absorption in the K\,{\sc i} lines
is observed to be considerably weaker than in the Na\,D lines.

Betelgeuse has been subjected to numerous investigations of its mass loss
rate using a variety of spectroscopic indicators. The K\,{\sc i}-based rate
may be compared with a couple of determinations using tracers of a major
constituent of the shell. Bowers \& Knapp (1987)  detected 21 cm H\,{\sc i}
emission to derive $\dot{M} \sim 1.1 \times 10^{-6}$ $M_\odot$ y$^{-1}$
 if $d$ = 140 pc.
Huggins et al. (1994) detected the C\,{\sc i} 609 $\mu$m fine structure
line and inferred a mass loss rate also of $\dot{M} \sim 1 \times 10^{-6}$
$M_\odot$ y$^{-1}$ for $d$ = 140 pc. The close agreement between these
results and ours is encouraging but must be largely fortuitous.
More recently, Harper et al. (2001) estimated $\dot{M} = 3.1 \pm 1.3
\times 10^{-6} M_\odot$ y$^{-1}$ from absorption features in ultraviolet
Fe\,{\sc ii} lines.

\subsection{Departures from Spherical Symmetry}

Azimuthal averaging produces a smooth run of the emission line's flux with
radial distance but masks a fascinating structure of incomplete shells and
clumps. This structure is illustrated in 
Fig.~\ref{fig2}, \ref{fig3}, and \ref{fig4}
 and additionally
in Fig.~\ref{fig5a}, and \ref{fig5b}. Several features of the structure are
seen by inspection: the appearance of thin incomplete shells, intense clumps
of small spatial extent and velocity dispersion, and lines of sight -
particularly far from the star and inside the outermost detected shell -
from which no 7699\AA\ emission is detected, and a seemingly sharp
boundary to the outermost shell. 

The following relations are helpful in discussing the properties of the
circumstellar shell.
At the assumed distance of 140 pc, 1 second of arc corresponds to a
projected distance of 140 AU or a light travel time of about 19 hours.
 The stellar angular diameter is taken to
be 0.044  arcsec (Dyck et al. 1998) or a stellar
radius of 3.1 AU. At an expansion velocity of 15 km s$^{-1}$, gas
travels 3.2 AU in 1 year or an angular distance of just 0.02  arcsec.

\subsubsection{Temporal Dependence?}

Our observations were obtained over several months in 1994. In different
observing runs, several successful attempts were made to acquire
a series of spectra
with the slit set relative to the
star at the same positions and orientations. A goal of these
observations was to search for changes in the velocity-position
maps.

Expected time scales
for changes to occur in the wind itself are much longer than a few months 
(see below). More rapid 
changes may result
from the rise and fall of bright (or dark) spots on the
stellar surface. High angular
resolution images of Betelgeuse's disk show a few (1 - 3)  bright spots
contributing  10 - 20 per cent of the light and changing on a timescale
of weeks to months (Wilson, Dhillon, \& Haniff 1997). With a light
travel time of about a month to a radial distance of about 30$^"$, intensity
fluctuations of (say) 20 \% evolving on a time scale of a month are not
going to produce dramatic changes in the fluorescent emission. Were a 
spectacularly bright spot to occur, a large intensity contrast would
be seen in the hemisphere facing the spot.  The appearance of shells and
especially the presence of clumps cannot be traced to spots on the
illuminating photosphere.
 
Changes in the density of potassium atoms depend on the rates of
ionization and recombination. With rates taken from Glassgold \& Huggins (1986),
we find the timescale for photoionization by the stellar radiation field to
be $t_{photo} \sim 0.1 \theta^2$ yr where $\theta$ is the distance from
the star expressed in  arcsec. The recombination timescale is
necessarily much longer: $t_{rec} \sim 30\theta^2$ yr (or $\theta^{1.3}$ with
the temperature dependence of the recombination coefficient 
taken into account). Except perhaps very
close to the star, these estimates show that changes
in K\,{\sc i} emission  occurring on a time scale of a few months are unlikely
to be traceable to variations in the photoionizations
by the stellar radiation field. 

Fig.~\ref{fig3} shows pairs of velocity-position maps taken at the
same slit position relative to the star but at 8 months apart. By inspection,
there is considerable similarity between the March and November observations 
for which the impact parameter of the slit was 13 
arcsec.   Gross structures may be deemed identical.
Smaller structures (here dubbed `clumps') show no discernible change in
location. Differences in shape of the
smaller structures may be attributed to differences in the
seeing: the seeing was about 1.2  arcsec for the March observation but
about 3  arcsec in November.

\begin{figure}
\centerline{\epsfig{figure=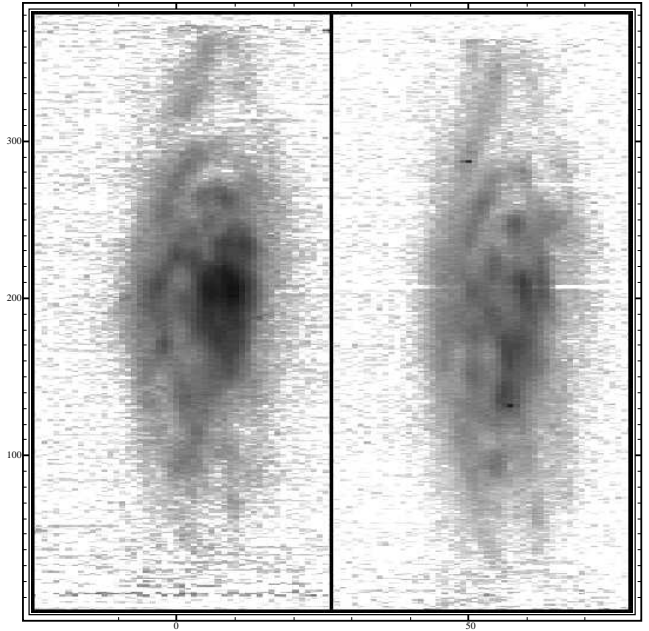,bbllx=29pt,bblly=170pt,bburx=552pt,bbury=697pt,clip=,width=8.5cm}}
\caption{Two observations at the same orientation
and slit position (21"
away from the star). The star corresponds to line  191.
Exposure time 30 min. Left panel: March 23, 1994 with 1.7" seeing. Right panel:
November 26, 1994 with seeing around 1.2". The general appearance
 of the 2 frames
is similar, but more differences are visible than on Fig.~\ref{fig3}.
The central condensation around pixel (5,210) is more conspicuous
in the left-hand frame, and differences in the clump structure
are obvious in the region around (5,270).}
\label{fig5a}
\end{figure}

\begin{figure}
\centerline{\epsfig{figure=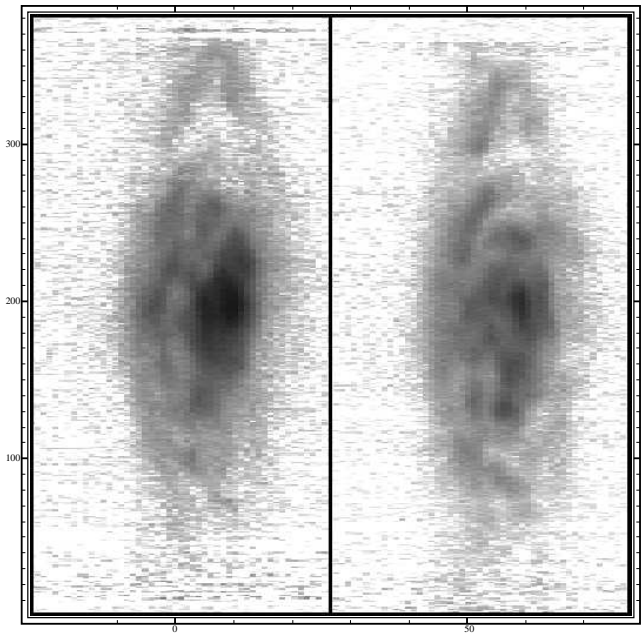,bbllx=29pt,bblly=170pt,bburx=552pt,bbury=697pt,clip=,width=8.5cm}}
\caption{Same as Fig.~\ref{fig5a}. These spectra were acquired 
directly following the ones of Fig.~\ref{fig5a}, at the same slit position.
While there is considerable similarity between this and Fig.~\ref{fig5a}, there are
some differences. These illustrate the changes in appearance
revealed by the rotation of the slit on the sky. 
Comparison of the right-hand panels of the two figures
shows differences in the regions around 
pixels (55,80), (55,250), and (50, 330).}
 \label{fig5b}
\end{figure}

Fig.~\ref{fig5a} and \ref{fig5b}
 show  similar pairs of velocity-position maps from the March
and November runs but with an impact
parameter of 21  arcsec.
The exposure in Fig.~\ref{fig5b} was obtained immediately following that in 
Fig.~\ref{fig5a}.
In these figures, there is evidence for changes in the smaller
structures. Indeed, the March pair of maps clearly differ in the inner parts
from the same area in the November maps. Close inspection shows
a few differences for clumps in successive maps in March and in November.
This is not surprising given that 
 successive maps owing to image rotation at the
coud\'{e} focus do not sample exactly the same portion of the
shell.  The telescope was guided on the star and the slit rotates
about the star. Changes in position on the sky recorded by a particular
pixel can exceed the angular size of some clumps, and, hence, these
may appear clearly in one image but not in the next. This instrumental
effect increases with increasing impact parameter which may explain in part why
changes are more evident in Fig.~\ref{fig5a} and \ref{fig5b} than in Fig.~\ref{fig3}. 

\subsubsection{A Distant Spherical Shell}\label{detached}

A striking feature of the velocity-position maps with impact parameters
greater than about 20  arcsec is the appearance of an elliptical
shape for the outermost emission (Fig.~\ref{fig4}.
Emission is very weak, possibly absent, beyond about 1 minute of arc
from the star.
 Ellipticity of isophotes in the
velocity-position map corresponds to a spherically symmetric 
expanding shell. Geometry of the slit position (at 33 arcsec from the star)
and the semi-major axis of the
ellipse show that the shell radius is 50 $\pm$ 2  arcsec. 
The maximum velocity width, the ellipse's semiminor axis,  
corresponds to an expansion (radial) velocity of  18 $\pm$ 2km s$^{-1}$. 
Note that the shell is clearly seen in Fig.~\ref{fig6} as the plateau 
followed by an intensity drop at about 50  arcsec. At that distance from
the star, the mean K\,{\sc i} density is rapidly decreasing with distance,
due to the interstellar UV field (see Fig.~\ref{fig6}
and Sec.~\ref{radial}), and the shell represents a considerable density
enhancement.

The thickness of the shell is 1 to 2   arcsec in position
on the sky and is unresolved in velocity, i.e. less than 2.6 km s$^{-1}$.
 Non-uniform emission over the spherical shell could result from
one or more of several factors: shell ejection occurred preferentially over
one hemisphere of the star; the excitation of the shell is non-uniform
owing to a bright spot on the stellar disk; the ionisation of potassium atoms is
greater over one hemisphere due to a chromospheric enhancement;
 the shell is plowing into
the irregular local interstellar medium and deceleration enhances the local
density non-uniformly.  

Circumstellar absorption imposed upon the photospheric spectrum  is
composed of two components in atomic (e.g., K\,{\sc i} 7699 \AA\ as
shown by Bernat \& Lambert [1975] and Goldberg et al. [1975]) and
the CO lines by  Bernat et al. (1979)  who
labelled the components S1 and S2 where the former has a
higher excitation temperature (200K vs 70K),  substantially larger
CO and K\,{\sc i} column densities
and  lower expansion velocities. The expansion velocity deduced from the
S2 absorption lines (20.2 km s$^{-1}$ from Goldberg et al., and 13 km
s$^{-1}$ from Bernat et al.) are consistent with our measurement. Note that
the absorption line measurements are made relative to photospheric lines
but the photospheric velocity is variable. Our measurement is a true
expansion velocity, relative to the center-of-mass of the star,
 independent of the stellar velocity, and of photospheric 
velocity variations.

Another notable difference between the S1 and S2 components is the very
narrow line width of the latter: a Doppler width of 1 km s$^{-1}$ for S2 but
4 km s$^{-1}$ for S1 from visual lines (Bernat 1977). The Doppler width
for S2 corresponds to a full width at half intensity of 1.7 km s$^{-1}$,
an estimate consistent with the upper limit obtained 
from our observation of the shell emission. Bernat et al.
from a comparison of the collisional and radiative rates for CO
excitation inferred that the former dominate and, hence, their
excitation temperature was approximately equivalent to the
gas kinetic temperature. Reference to a dust shell model (Tsuji 1979)
showed a gas temperature of 70 K was achieved at about 55 arcsec from
the star, a prediction quite consistent with our observed shell
radius.

Judged by  velocity of expansion,  line width, and 
shell radius as  measured or estimated from the CO lines and our 7699\AA\
velocity-position maps, we identify the circumstellar gas contributing
the  S2 absorption lines with the outermost
thin shell seen in Fig.~\ref{fig4}. This identification 
allows interesting further deductions to be made.  In particular, the
volume density of CO molecules and neutral K atoms may be estimated
from their column density and the shell thickness. The equivalent
width of the K\,{\sc i} 7699\AA\  S2 line estimated from Goldberg et al.
(1975, their Figure 1a) gives a column density N(K\,{\sc i}) $\simeq
1.3 \times 10^{11}$ cm$^{-2}$ on the assumption that the line is unsaturated. 
After correction for seeing, the shell's thickness may be about 1 second of
arc. 
At $d$ = 140 pc, a shell thickness of 1 second of arc 
corresponds to 2 $\times 10^{15}$ cm. Then, the K\,{\sc i}
density in the shell 
is about 6 $\times 10^{-5}$ cm$^{-3}$. In contrast, Rodgers \& Glassgold's
 (1991) model predicted a density of $n$(K\,{\sc i})$\simeq 2
 \times 10^{-6}$ cm$^{-3}$ at the shell's distance.
This difference is not surprising as the shell by inspection
of Fig.~\ref{fig4} is evidently a considerable
density enhancement above the gas at smaller and greater distances
In addition, the intensity variation around the shell is obvious
implying that the cut through the shell to the star that gives the
K\,{\sc i} absorption line may not be representative of mean
conditions in the shell.

If the shell at 55 arcsec is identified with Bernat et al.'s S2 shell, their
measured column density (1.2 $\times 10^{16}$ cm$^2$) translates to
a volume density $n$(CO)$ \simeq
 6$ cm$^{-3}$.
The ratio of the CO and K\,{\sc i} column densities suggests a density
ratio
$n$(CO)/$n$(K\,{\sc i})$
\sim 9 \times 10^4$. The elemental abundance ratio in the
photosphere is probably a little less than the solar ratio of 2700 on
account of a lower C abundance resulting from the first dredge-up. If
C has been reduced by about 50\%, and K is not depleted onto grains in the
wind, the expected elemental ratio in the wind is 1350. The  higher ratio
in the S2 shell implies, as expected, that K is largely ionized:
the fraction of neutral  K atoms is $y_{\rm K} \simeq 0.014$
 provided that most of the
C is in CO.
Huggins et al.'s (1994) prediction of ionization of C and association into
CO through the shell suggests CO accounts for about 20 \% of the
carbon. Then, $y_{\rm K} \sim$ 0.003. 
Rodgers \& Glassgold (1991) `standard' model predicts $y_{\rm K} \simeq 0.01$
at the radial distance equivalent to 50  arcsec at $d$= 140 pc. 
This is fair agreement with observation given the various (uncertain)
ingredients that enter into both estimates. The mass in the shell amounts to
about 10$^{-2}$~M$_{\odot}$, at a distance of 140 pc and assuming 
$y_{\rm K} = 0.01$.

As noted above the S2 shell is not a consequence of a varying photospheric
illumination; the light travel time across the shell is only about a day,
and the shell is only about 50 light days from the star. In addition,
the shell seen in absorption in the 1970s (Goldberg et al. 1975; Bernat et al.
1979) is plausibly identified with the K\,{\sc i} emission shell seen in
1994. The simplest explanation for the shell may be that it is mass ejected
at a substantially higher than average rate at an earlier time. If the
expansion velocity has been constant, gas travelling at 18 km s$^{-1}$ and
now at 50  arcsec from the star was ejected 1900 years ago ($d$ = 140 pc
is assumed). A shell thickness of 1 second of arc implies ejection
lasted about 40 yr but this is a guess because the velocity of ejection
presumably varied too. The shell as it appears in Fig.~\ref{fig4} is not
uniformly bright. It is unlikely that the observed contrast around the
shell is due to
the presence of bright spot  on the disk at the time of the observations. 
A possible explanation is that mass ejection occurred preferentially
above a spot 1900 years ago. 
Stellar rotation (the rotation period of Betelgeuse is at least 
a few years), 
could have led to  material being ejected into the
shell at a variable rate over the lifetime of the spot.
There appear to be at least two other shells of moderate completeness which
suggests that significant increases in mass loss may occur every
several hundred years. Multiple shells were also detected around 
$\mu$ Cep by Mauron (1997), with a characteristic time scale between two
successive increases in mass loss of about one thousand years.
Detached shells are also observed around AGB stars, where they are
attributed to a
Helium shell flash that induces both a mass loss increase and a
two-wind interaction due to the increased outflow velocity 
(Steffen and Sch\"onberner 2000). Clearly, an alternative mechanism
must be operating in supergiant stars like Betelgeuse.
 
Future observations of the K\,{\sc i} emission from the shell should be
made to search for changes in the S2 shell. Measuring the S2's expansion
will be a challenge: the proper motion expected  is a mere 0.03 
arcsec per year.

\subsubsection{Clumps}

A striking feature of our velocity-position maps, especially those acquired
in the best seeing is the appearance of clumps or knots of emission;
Fig.~\ref{fig3}, \ref{fig5a} and \ref{fig5b} show clumps extending in size down to
 the velocity and spatial
resolution of the map.  There are locations where clumps appear
connected suggesting they are condensations in a sheet or shell. 
In general, the collection of clumps in many velocity-position
maps accounts for more than half of the emitted flux. The impression 
of elliptical structures is largely gained from gas at large
angular distances from the star. Many of our exposures, especially those
taken with small impact parameters, were too short to reveal faint
distant structures.

These are the first observations of such clumps in Betelgeuse's
wind.  Mauron (1997) from long-slit spectra showing
K\,{\sc i} emission from $\mu$~Cephei reports `inhomogeneities' with
an apparent size of 1$^{\prime\prime}$ to 3$^{\prime\prime}$ when the
angular resolution was about 0.6$^{\prime\prime}$. In velocity space,
$\mu$~Cep's clumps were unresolved at the moderate resolution of 40 km s$^{-1}$
of these observations. Since $\mu$~Cep is about 5 times more distant than
Betelgeuse, its smallest clumps would appear to be 5 to 10 larger than
those we have detected.

There is a hint that the prevalence of clumps is higher close to the
star than at the largest distances probed by our spectra. 
This encourages the speculation that
the clumps' origin may be traceable through the inhomogeneous
inner shell (Lim et al. 1998) and chromosphere to the stellar surface.
The greater visibility of shells at large distances may be due to
the dissolution and merger of clumps at a common distance. 
Although clumps are seen by fluorescent emission, their shape and size
are attributable to an enhanced concentration of potassium atoms in the wind
and not to intensity variations over the surface of the illuminating
photosphere; a bright photospheric spot will enhance fluorescence over
approximately 2$\pi$ of solid angle not a mere few  arcsec.
The clumps are probably not regions of reduced ionization of
potassium atoms because the characteristic timescale for photoionization 
is shorter that the expansion timescale (the time required to reach a given
angular distance at a constant expansion velocity): it is a factor of 100
faster close to the star and about a factor of 10 at 50 
arcsec from the star.

Supposing the clumps represent coherent structures related to the
process of mass loss, a change of clump
characteristics with radial distance provides information about the clumps'
internal structure. Consider  a clump moving radially
outwards at velocity $v_{exp}$ with an internal velocity dispersion $v_{disp}$.
Its radial size $\phi$ increases linearly with radial distance out from the
star $\Phi$, i.e., $\Delta\phi$/$\Delta\Phi$ = $v_{disp}/v_{exp}$. 
Identifying $v_{disp}$ with the sound speed, we estimate
$v_{disp}$ = 1 km s$^{-1}$. Since $v_{exp} \simeq 15 $
km s$^{-1}$, a clump of about 2  arcsec at 10  arcsec from
the star would expand to about 5  arcsec at a radial distance of
50  arcsec. In the inner shell, the inter-clump distances are a
few  arcsec. Then, if dispersion of the clump
 perpendicular to the radial flow occurs at about 1 km s$^{-1}$, the clumps
at a common distance may well merge after a few tens of  arcsec.

 Our observations are consistent with this result
in that the maximum velocity extent of a  velocity-position
map  is not greatly different for small and large impact parameters.

\section{Concluding Remarks}

Our study of the K\,{\sc i} emission from Betelgeuse's circumstellar
shell reveals new aspects of the shell and its potassium atoms. Spherical
symmetry is no more than a rough approximation to the distribution of
potassium atoms but the radial distribution of the brightness closely
follows the predictions for a model proposed by Rodgers \& Glassgold (1991)
implying a mass loss rate of about 2 $\times 10^{-6}$  $M_\odot$ y$^{-1}$.
But the striking result of our study is the presence of clumps of
potassium atoms with the smallest clumps being
unresolved spatially and kinematically. The clumps likely originate from
cool structures close to the star. Observations suggest that they disappear
at large distances from the star.

Observations at a site of excellent seeing are to be sought in order to
trace the clumps back closer to the star. Other supergiants and giants
are observable in both Na\,D and K\,{\sc i} emission, as published
reports demonstrate, but the occurrence of clumps in the winds from 
such stars is as yet unknown.  To explore the innermost regions of 
these shells will require the spatial resolution of a space,  
or a diffraction limited ground-based telescope.
Betelgeuse's chromosphere has already been explored in the ultraviolet continuum
and the Mg\,{\sc ii} h and k lines (Uitenbroek, Dupree, \& Gilliland 1998).
Extension of this work to other lines, and other stars would be rewarding.

\begin{acknowledgements}
We thank N. Mauron and E. Josselin for inspiring discussions of the Betelgeuse shell.
\end{acknowledgements}

\end{document}